\documentclass[12 pt, amsfonts,amsmath, amssymb,color]{article}
\usepackage{authblk}
\def\baselinestretch{1.0}
\usepackage{amsmath,amssymb,amsfonts,latexsym,float,graphics,epsfig}
\usepackage{subfig}
\usepackage{verbatim}
\usepackage{relsize}
\usepackage{hyperref}
\usepackage{graphicx}
\usepackage{bm}
\usepackage{epstopdf}
\usepackage{slashed}
\usepackage{color}

\def\be{\begin{equation}}
\def\ee{\end{equation}}
\def\bea{\begin{eqnarray}}
\def\eea{\end{eqnarray}}

\begin{document}

\renewcommand\theequation{\arabic{section}.\arabic{equation}}
\catcode`@=11 \@addtoreset{equation}{section}
\newtheorem{axiom}{Definition}[section]
\newtheorem{theorem}{Theorem}[section]
\newtheorem{axiom2}{Example}[section]
\newtheorem{lem}{Lemma}[section]
\newtheorem{prop}{Proposition}[section]
\newtheorem{cor}{Corollary}[section]

\newcommand{\ben}{\begin{equation*}}
\newcommand{\een}{\end{equation*}}

\let\endtitlepage\relax

\begin{titlepage}
\begin{center}
\renewcommand{\baselinestretch}{1.5}  

\vspace*{-0.5cm}

{\fontsize{19pt}{22pt}\bf{Novel logarithmic corrections to black hole entropy}}

\vspace{9mm}
\renewcommand{\baselinestretch}{1}  

\centerline{\large{Aritra Ghosh$^{\ddagger}$}\footnote{ag34@iitbbs.ac.in}, \large{Sudipta Mukherji$^{\dagger}$}\footnote{mukherji@iopb.res.in} and \large{Chandrasekhar Bhamidipati$^\ddagger$}\footnote{chandrasekhar@iitbbs.ac.in}}

\vspace{5mm}
\normalsize
\textit{$^\ddagger$School of Basic Sciences, Indian Institute of Technology Bhubaneswar, Odisha, 752050, India}\\
\textit{$^\dagger$Institute of Physics, Sachivalaya Marg, Bhubaneswar, 751005, India}\\
\textit{$^\dagger$Homi Bhabha National Institute, Training School Complex, Anushakti Nagar, Mumbai, 400085, India}\\
\vspace{5mm}

\begin{abstract}
For a thermodynamic system, apart from thermal fluctuations, there are also fluctuations in thermodynamic volume when the system is in contact with a volume reservoir. For the case of black holes in anti-de Sitter spacetimes, the effect of thermal fluctuations on the entropy is well studied. The aim of this work is to compute novel logarithmic corrections to black hole entropy coming from simultaneous fluctuations of energy and thermodynamic volume. We work in the isothermal-isobaric ensemble and first obtain a general form of corrections to entropy which are valid for any thermodynamic system.  Applying the formalism to Kerr black holes in AdS reveals that the black hole entropy gets corrected as: $\mathcal{S} = S_0 - k \ln S_0 + \cdots$ where $S_0$ is given by the Bekenstein-Hawking formula and $k = - 1$. The same leading coefficient is also obtained in the canonical ensemble, i.e. by considering energy fluctuations alone. This coefficient is found to be unaltered in the slowly rotating and high temperature limits. 
\end{abstract}
\end{center}
\vspace*{0cm}


\end{titlepage}
\vspace*{0cm}

\section{Introduction}

Finding the nature of microscopic degrees of freedom of black holes is one of the major goals of quantum gravity and the indispensable tool is the Boltzmann entropy formula,
\begin{equation} \label{entropy}
S = k_B \, \ln \, \Omega_m\,
\end{equation}
where $k_B$ is the Boltzmann constant\footnote{In the rest of this paper, we will set \(k_B = 1\).} and $\Omega_m$ is the number of microstates of the system. A key test of any quantum theory of gravity rests in a successful matching of microscopic and macroscopic calculations of black hole entropy \cite{Bekenstein:1973ur}-\cite{Hawking:1976de}. In this direction, several different methods have been employed for computation of entropy via microscopic counting of states in extremal black holes and general situations~\cite{Vafa}-\cite{Mathur}. Remarkable progress has been made in understanding microscopic aspects in a holographic setting \cite{Maldacena:1997re}-\cite{Witten:1998zw}. For large black holes, the Bekenstein-Hawking entropy gets logarithmic corrections \cite{9407001}-\cite{Pourhassan:2017qxi}, which have been computed for extremal and non-extremal black holes, using Euclidean quantum gravity methods~\cite{1005.3044}-\cite{9412020} as well as loop quantum gravity and the saddle point method \cite{9801080}-\cite{open}. Black hole thermodynamics has been investigated in a canonical ensemble (or a grand canonical ensemble if there are other conserved charges which can fluctuate). However, for asymptotically flat black holes, a thermal equilibrium cannot be achieved. This is so because the black hole (subsystem) either accretes matter or radiates without limit depending on the whether the surrounding (reservoir) temperature is higher or lower than the Hawking temperature. Although, the situation improves in the presence of a negative cosmological constant, one still needs to restrict to certain parameter range for the canonical ensemble to be well defined for AdS black holes. \\

\noindent
Typically, the energy of a system in contact with a thermal reservoir fluctuates about its equilibrium value due to continuous random transitions among its microstates. For non-extremal black holes in AdS, it is now well known that such energy fluctuations lead to logarithmic entropy corrections of the form \cite{0002040},
\begin{equation}\label{k}
\mathcal{S} = S_0 - k \ln S_0 + \cdots \,
\end{equation}
where $S_0$ is the standard Bekenstein-Hawking term and $k$ is a number which varies for different black holes. Such logarithmic corrections have been computed using several different methods such as Euclidean partition function, canonical ensemble and AdS/CFT, though not all agree (see~\cite{Sen:2012dw} for a nice summary). \\

\noindent
For any thermodynamic system with the notion of volume, just like fluctuations of energy, there are also fluctuations of thermodynamic volume when the system is in contact with a volume reservoir. These fluctuations are expected to correct the equilibrium value of the entropy as well. Hitherto, the effect of the latter set of fluctuations on the thermodynamics of black holes has not been investigated yet. In this paper, we show that within the framework of extended thermodynamics\footnote{See also \cite{Henneaux:1984ji}-\cite{K}.} \cite{Kastor:2009wy}, the Bekenstein-Hawking entropy of a black hole receives novel logarithmic corrections from simultaneous fluctuations of energy and thermodynamic volume since the system can be taken to be in contact with both energy and volume reservoirs. These corrections to the black hole entropy are distinct from those found in earlier studies in the sense that the earlier results \cite{0002040} cannot be obtained by taking a limit of vanishing volume fluctuations. Generally, black holes placed in a cavity with reflecting boundary conditions or in an AdS spacetime can be treated within the canonical ensemble \cite{Hawking:1982dh}-\cite{Chamblin:1999hg} where the Hawking temperature \(T\) at thermodynamic equilibrium can be thought of as being attained through equilibration with a thermostat at the same temperature. In the extended thermodynamics framework, one has the novel concept of pressure arising due to a dynamical cosmological constant. It may then be quite natural to think of AdS as being a barostat with the cosmological constant \(\Lambda\) leading to a pressure \(P\) as,
\begin{equation}\label{P}
P = - \frac{\Lambda}{8\pi}\,= \frac{(d-1)(d-2)}{16\pi l^2}
\end{equation}
where $\Lambda$ is the cosmological constant, $l$ denotes the AdS radius and $d$ is the number of spacetime dimensions. The conjugate thermodynamic volume is defined as \(V = (\partial H/\partial P)_S\), where \(H(S,P) = E + PV = M\) is the enthalpy of the system and $M$ is the ADM mass of the black hole. Thermodynamic volume satisfies a novel reverse isoperimetric inequality~\cite{Cvetic:2010jb} and has recently been found to give a distinct approach to holographic complexity~\cite{Balushi:2020wkt}. Formally, the Euclidean action equates to the Gibbs free energy of the thermodynamic system (see for example \cite{Kubiznak:2012wp}) and therefore, extended black hole thermodynamics is equivalent to the thermodynamic limit of the isothermal-isobaric ensemble with the black hole equilibrating with both a thermostat at inverse temperature \(\beta\) and a barostat at pressure \(P\). This identification is consistent with taking the enthalpy \(H(S,P)\) as the key thermodynamic energy function \cite{Kastor:2009wy} and the Gibbs potential \(G(T,P)\) as the natural free energy \cite{Kubiznak:2012wp}.\\

\noindent
Given the above set up, in section-(\ref{MEF}), we discuss some aspects of thermodynamic fluctuations and their effect on the microcanonical entropy within the Gaussian approximation. In subsection-(\ref{TP}), we consider an arbitrary thermodynamic system (not necessarily a black hole) in the \((T,P)\)-ensemble and obtain an exact formula for the logarithmic entropy correction due to fluctuations in energy and volume. Following this, in section-(\ref{bh}), we apply this general formula to rotating black holes in anti-de Sitter spacetimes to obtain logarithmic corrections to the black hole entropy. The leading correction term within the first order is found to be of the form \(-k \ln S_0\) with \(k = -1\) in four dimensions. Quite remarkably, this coefficient remains unchanged in the high temperature and small angular momentum limits. Finally, we conclude the paper in section-(\ref{conclusions}). Explicit calculations of the expressions of the covariances are summarised in appendix-(\ref{appendixa}). Some remarks for spherically symmetric black holes are made in appendix-(\ref{appendixb}). 

\section{Microcanonical entropy and fluctuations}\label{MEF}
In this section, we will discuss thermodynamic fluctuations (see for example \cite{callen}) and their role in modifying the microcanonical entropy of a system up to the lowest order. Let us begin with the canonical ensemble where the system is in contact with a thermostat and the volume is fixed by the boundaries enclosing the system. In subsection-(\ref{TP}), we shall consider the effect of a barostat fixing the pressure (rather than volume) of the system at the boundary. \\

\noindent
The system is in contact with a thermostat at temperature \(T = 1/\beta\). Recall that the canonical partition function is readily defined as the Laplace transform of the density of states \(\rho(E,V)\) as,
\begin{equation}
  Z(\beta,V) = \int_{0}^{\infty} \rho(E,V) e^{-\beta E} dE.
\end{equation} This means that the density of states can be obtained as,
\begin{equation}\label{xxx}
  \rho(E,V) = \frac{1}{2 \pi i}\int_{\gamma - i\infty}^{\gamma + i\infty} Z(\beta,V)e^{\beta E} d\beta = \frac{1}{2 \pi i}\int_{\gamma - i\infty}^{\gamma + i\infty} e^S d\beta
\end{equation}
where, one has used \(S = \beta E - \ln Z\). At this stage, we shall incorporate the notion of thermal fluctuations. In the Gaussian approximation, we expand the entropy \(S=S(\beta)\) to the second order about the equilibrium point \(S_0\) as,
\begin{equation}
  S(\beta) \approx S_0 + \frac{\partial^2 S_0}{\partial \beta^2} (\beta - \beta_0)^2.
\end{equation}
Here, it is to be understood that, 
\begin{equation}
\frac{\partial^2 S_0}{\partial \beta^2} = \frac{\partial^2 S(\beta)}{\partial \beta^2}\bigg|_{\beta = \beta_0}.
\end{equation}
 Noting that in the canonical ensemble, one has,
 \begin{equation}
   \frac{\partial^2 S_0}{\partial \beta^2} =  (\Delta E)^2
 \end{equation} we can write, \(S(\beta) \approx S_0 + (\Delta E)^2 (\beta - \beta_0)^2\). Substituting this into eqn (\ref{xxx}) and performing the integral one finds,
 \begin{equation}
   \rho(E,V) = \frac{e^{S_0}}{\sqrt{2 \pi (\Delta E)^2 }}.
 \end{equation}
Thus, the microcanonical entropy which is defined as \(\mathcal{S} := \ln \rho(E,V)\) acquires a logarithmic correction given by,
\begin{equation}\label{correctedentropycanonical}
  \mathcal{S} = S_0 - \frac{1}{2} \ln  (\Delta E)^2  + ({\rm higher~order~terms}).
\end{equation}
One should note that \((\Delta E)^2\) is the variance in internal energy of the system due to fluctuations about equilibrium. It is positive definite ensuring that the entropy corrections are well defined. It is a simple exercise to show that \((\Delta E)^2 = T^2 C_V\) and therefore the condition for stability against fluctuations is \(C_V > 0\). For a typical extensive system consisting of \(N\) degrees of freedom, one can write \(C_V = N c_V\) and \(E = N \epsilon\) where \(c_V\) and \(\epsilon\) are the relevant quantities per particle. Then, one gets,
\begin{equation}
\frac{\Delta E}{E} \sim \frac{1}{\sqrt{N}}
\end{equation} meaning that relative energy fluctuations become negligible in the thermodynamic limit, i.e. \(N \rightarrow \infty\). \\

\noindent
Eqn (\ref{correctedentropycanonical}) was applied to black holes long back \cite{parthasarathi} and novel insights into the nature of black hole entropy were obtained in good agreement with quantum geometry (see for example \cite{0104010}) and string theory expectations \cite{Carlip,parthasarathi,Mukherji:2002de}. As such, these developments warrant similar explorations in extended phase space thermodynamics where the cosmological constant is promoted to the status of a full thermodynamic variable, namely the pressure. Let us point out an important difference between the setting considered in \cite{parthasarathi} and the present scenario. In \cite{parthasarathi}, the mass \(M\) of the black hole was interpreted as the energy of the spacetime, i.e. \(M = E\) and as such \((\Delta E)^2\) corresponds to fluctuations of the black hole mass (the black hole being the system under consideration). In extended thermodynamics however, the mass \(M\) of the black hole is identified with the enthalpy of the spacetime such that energy \(E = M - PV\) is defined by a Legendre transform. Thus, in the present case \((\Delta E)^2\) corresponds to the variance of the quantity \(E = M - PV\) rather than \(M\) itself. Therefore, our results in general differ from the earlier results reported in \cite{parthasarathi}. Below, we find the general form of logarithmic entropy corrections if the system interacts with a barostat (volume reservoir) in addition to the thermostat. This ensemble is motivated from extended thermodynamics since the cosmological constant appearing in the black hole solution leads to a pressure term [via eqn (\ref{P})] and the black hole mass contains \(P\) as an independent variable with \(V\) being a derived quantity. 

\subsection{Isothermal-isobaric ensemble} \label{TP}
The partition function of a system connected to both a thermostat at inverse temperature \(\beta\) and a barostat at pressure \(P\) is (see for example \cite{isob1,isob2}),
\begin{equation}\label{lt}
  \Delta(\beta,\beta P) = C \int_{0}^{\infty} Z(\beta,V) e^{-\beta PV} dV
\end{equation} where \(Z(\beta,V)\) is the canonical partition function. The pre-factor \(C\) is a constant with suitable dimensions necessary to make \(\Delta(\beta,\beta P)\) dimensionless. Now the Laplace transform of eqn (\ref{lt}) may be inverted to give,
\begin{equation}\label{11}
  Z(\beta,V) = \frac{C^{-1}}{2 \pi i} \int_{\gamma - i\infty}^{\gamma + i\infty} \Delta(\beta, \beta P) e^{\beta PV} d(\beta P).
\end{equation}
But the canonical partition function is defined as,
\begin{equation}
  Z(\beta,V) = \int_{0}^{\infty} \rho(E,V) e^{-\beta E} dE \,
  \end{equation} which can be inverted to give,
  \begin{equation}\label{2}
    \rho(E,V) = \frac{1}{2 \pi i} \int_{\delta - i\infty}^{\delta + i\infty} Z(\beta,V) e^{\beta E} d\beta.
  \end{equation}
  Combining eqns (\ref{11}) and (\ref{2}), one obtains,
  \begin{equation}
    \rho(E,V) = \frac{C^{-1}}{(2 \pi i)^2} \int_{\gamma - i\infty}^{\gamma + i\infty} \int_{\delta - i\infty}^{\delta + i\infty} \Delta(\beta,\beta P) e^{\beta (E + PV)} d\beta d(\beta P).
  \end{equation}
Note that \(\Delta(\beta,\beta P) = e^{-\beta G}\) where \(G = E + PV - TS\) is the Gibbs free energy. With this, one finally finds,
\begin{equation}\label{3}
  \rho(E,V) = \frac{C^{-1}}{(2 \pi i)^2} \int_{\gamma - i\infty}^{\gamma + i\infty} \int_{\delta - i\infty}^{\delta + i\infty} e^S d\beta d(\beta P)
\end{equation} where \(S\) is the entropy at an arbitrary temperature and pressure, not just at equilibrium. Consequently, one can generically express  \(S = S(\beta,\beta P)\). In order to take into account the effect of fluctuations up to the lowest order, we expand the entropy about its equilibrium value \(S_0\) as,
\begin{equation}\label{Sexpand}
  S(\beta,\beta P) \approx S_0 + \frac{1}{2}\bigg(\frac{\partial^2 S_0}{\partial \beta^2}(\beta - \beta_0)^2 + 2\frac{\partial^2 S_0}{\partial \beta \partial (\beta P)}(\beta - \beta_0)(\beta P - (\beta P)_0)  + \frac{\partial^2 S_0}{\partial (\beta P)^2}(\beta P - (\beta P)_0)^2 \bigg) . \,
\end{equation} Substituting this into eqn (\ref{3}) and performing the integrals, one finds,
\begin{equation}
   \rho(E,V) = \frac{C^{-1}}{(2 \pi )} \frac{e^{S_0}}{\sqrt{D}}
\end{equation}
where,
\begin{equation} \label{D}
   D = \partial^2_\beta S_0\,\partial^2_{\beta P} S_0- \left(\partial_\beta \partial_{\beta P} S_0 \right)^2\, .
\end{equation}
Here it should be noted that, 
\begin{equation}
\frac{\partial^2 S_0}{\partial x^i \partial x^j} = \frac{\partial^2 S}{\partial x^i \partial x^j}\bigg|_{\beta = \beta_0, \beta P = (\beta P)_0}
\end{equation} where \(x^i = \beta, \beta P\). In other words, the derivatives are evaluated at equilibrium. For consistency, one requires $D>0$. Upon ignoring the constants, the microcanonical entropy \(\mathcal{S} := \ln \rho(E,V)\) is,
\begin{equation}\label{entropycorrections}
  \mathcal{S} = S_0 - \frac{1}{2} \ln D + ({\rm higher~order~terms}).
\end{equation} This is the general expression for the corrected microcanonical entropy incorporating the lowest order effect of fluctuations about thermodynamic equilibrium for a system in the isothermal-isobaric ensemble. In the following subsection, we show that \(D\) defined in eqn (\ref{D}) is the determinant of the covariance matrix associated with thermodynamic fluctuations. Thus, positivity of \(D\) is necessary for stability against fluctuations.

\subsubsection{Covariance matrix and stability}
 Let us begin by noting that \(S = \ln \Delta(\beta,\beta P) + \beta(E + PV)\) which means we can write,
\begin{equation}
 \frac{\partial^2 S}{\partial x^i \partial x^j} = \frac{\partial^2 \ln \Delta}{\partial x^i \partial x^j}  \end{equation} where \(i,j = 1,2\) with \(x^1 = \beta\) and \(x^2 = \beta P\). This implies,
 \begin{equation}
   \partial^2_\beta S_0 = \bigg(\frac{\partial^2 \ln \Delta(\beta,\beta P)}{\partial \beta^2}\bigg)_{\beta_0,(\beta P)_0}  = (\Delta E)^2
 \end{equation} and,
 \begin{equation}
   \partial^2_{\beta P} S_0 = \bigg(\frac{\partial^2 \ln \Delta(\beta,\beta P)}{\partial (\beta P)^2}\bigg)_{\beta_0, (\beta P)_0}   = (\Delta V)^2 .
 \end{equation} Similarly,
 \begin{equation}
   \partial_\beta\partial_{\beta P} S_0 =  \langle \Delta E \Delta V \rangle .
 \end{equation} We reiterate that the derivatives are all evaluated at thermodynamic equilibrium. Thus, \(D\) is the determinant of the covariance matrix. It should be noted that \(\langle \Delta E \Delta V \rangle\) is the covariance of energy and thermodynamic volume. This is not the same as \((\Delta E) (\Delta V)\) which is the product of their standard deviations. Using the above, $D$ in eqn (\ref{D}) is,
\begin{equation}\label{entropycorrections1}
  D =  (\Delta E)^2 \,  (\Delta V)^2 - \langle \Delta E \Delta V \rangle ^2.
\end{equation}
Thus, the entropy of a general system (not necessarily a black hole) in contact with an energy and a volume reservoir gets corrected up to the lowest order in a manner summarised in eqn (\ref{entropycorrections}). Computing the exact expressions for \( (\Delta E)^2 \), \( (\Delta V)^2 \) and \(\langle \Delta E \Delta V \rangle\) we get (see appendix-(\ref{appendixa})),
\begin{eqnarray}
    (\Delta E)^2 &=& T^2 C_P - 2P V T^2 \alpha_V + P^2 TV \kappa_T,  \label{EE} \\
     (\Delta V)^2 &=& TV \kappa_T, \label{VV} \\
     \langle \Delta E \Delta V \rangle &=& T^2 V \alpha_V - PT V \kappa_T.  \label{EV}
  \end{eqnarray}
Here, \(C_P\), \(\alpha_V\) and \(\kappa_T\) respectively are the specific heat at constant pressure, the volume expansivity and the isothermal compressibility, defined as,
\begin{equation}
  C_P= \bigg(\frac{\partial H}{\partial T}\bigg)_P, \hspace{5mm} \alpha_V = \frac{1}{V}\bigg(\frac{\partial V}{\partial T}\bigg)_P, \hspace{5mm} \kappa_T = -\frac{1}{V}\bigg(\frac{\partial V}{\partial P}\bigg)_T.
\end{equation}
These covariances are proportional to the number of degrees of freedom in the system. For a typical extensive system with \(N\) degrees of freedom, one can introduce specific volume \(v = V/N\) such that eqns (\ref{EE})-(\ref{EV}) read, 
\begin{eqnarray}
    (\Delta E)^2 &=& N( T^2 c_P - 2P v T^2 \alpha_V + P^2 T v \kappa_T),  \label{EE1} \\
     (\Delta V)^2 &=& N T v \kappa_T, \label{VV1} \\
     \langle \Delta E \Delta V \rangle &=& N(T^2 v \alpha_V - PT v \kappa_T)  \label{EV1}
  \end{eqnarray}
where \(C_P = N c_P\) and the response functions \(\alpha_V\) and \(\kappa_T\) are already independent of system size. Thus, the relative fluctuations in energy and volume scale with \(N\) as,
\begin{equation}\label{evn}
 \frac{\Delta E}{E} \sim \frac{1}{\sqrt{N}}, \hspace{5mm}  \frac{\Delta V}{V} \sim \frac{1}{\sqrt{N}}
\end{equation} and become negligible in the thermodynamic limit, i.e. \(N \rightarrow \infty\). \\

\noindent
Now, upon substituting eqns (\ref{EE}), (\ref{VV}) and (\ref{EV}) into eqn (\ref{entropycorrections1}), one gets after some simplification,
\begin{equation}\label{D1}
  D = T^3 V C_V \kappa_T
\end{equation}
where we have used the following general relationship between the response functions,
\begin{equation}
C_P = C_V + \frac{T V \alpha_V^2}{\kappa_T}\, .
\end{equation}
Stability against thermodynamic fluctuations around equilibrium requires \(D > 0\) and this condition translates to,
\begin{equation}\label{condition}
  C_V \kappa_T > 0 \,
\end{equation} which must necessarily hold for the logarithmic corrections in the Gaussian approximation expressed in eqn (\ref{entropycorrections}) to be meaningful. Further, since fluctuations only up to the lowest order are being considered, the validity of our results is expected to get seriously compromised near the critical point where correlations and fluctuations are large, as will be seen in the following section.\\

\noindent
\subsubsection{Ideal gas and van der Waals fluid} As a simple example, consider an ideal gas whose equation of state reads \(PV = NT\). Thus, \(\kappa_T = V/NT\) and one gets from eqn (\ref{entropycorrections}) the leading logarithmic correction following expression,
\begin{equation}\label{ei}
  \mathcal{S} \approx S_0 - \ln (TV) \,
\end{equation} where we have disregarded the constant as well as higher order terms and have set \(C_V = 3N/2\) from the equipartition theorem. In the above expression, \(S_0\) is the entropy of the ideal gas being given by the Sackur-Tetrode equation. An identical calculation for the van der Waals fluid with equation of state $P = NT/(V-Nb)- aN^2/V^2$ gives the following result for corrected entropy after ignoring a constant term,
\begin{equation}
  \mathcal{S} \approx S_0 - \frac{1}{2} \ln \bigg(\frac{ T^3 V^3 (V-N b)^2}{T V^3 -2 a N (V-Nb)^2}\bigg)
\end{equation} where we have set \(C_V = 3N/2\). One should note that the positivity of the expression inside the logarithm is guaranteed by the positivity of \(\kappa_T\) which is indeed the case if we are away from the critical temperature. However, \(D\) diverges at the critical point (due to \(\kappa_T\) diverging) and therefore the logarithmic corrections break down. Further, below the critical temperature, there are some regions at which the compressibility is negative, rendering the corrections invalid. Thus, the logarithmic corrections are sensible away from the critical point where fluctuations are small and the Gaussian approximation works. \\

\noindent
It is also easy to check that one obtains eqn (\ref{ei}) in the limit \(a \rightarrow 0, b \rightarrow 0\). In either of the cases, we get non-trivial logarithmic corrections to the entropy due to fluctuations in energy and volume.

\section{Rotating black holes in AdS spacetimes} \label{bh}
Now that we have understood how logarithmic corrections to entropy arise due to fluctuations in volume in addition to those in the energy, we shall apply them to rotating (Kerr) black holes in AdS. We consider the case of Kerr-AdS black holes in four dimensions although a generalisation to higher dimensions is possible. The metric describing the Kerr-AdS solution has the following well known line element in Boyer-Lindquist coordinates,
\begin{equation}\label{1}
ds^{2}=-\frac{\chi}{\Omega}\bigg[dt-\frac{a\sin^2\theta}{k}
d\phi\bigg]^2 +\frac{\Omega}{\chi}dr^{2}+\frac{\Omega
}{\tilde{P}}d\theta^{2}+\frac{\tilde{P}\sin^{2}\theta}{\Omega
}\bigg[a dt-\frac{(r^2+a^2)}{k}d\phi\bigg]^2\, .
\end{equation}
Here,
\begin{eqnarray}
\chi=(r^2+a^2)\bigg(1+\frac{r^2}{l^2}\bigg)-2mr, \hspace{4mm} \Omega=r^2+a^2\cos^2\theta, \\
 k=1-\frac{a^2}{l^2}, \hspace{4mm}  \tilde{P}=1-\frac{a^2}{l^2}\cos^2\theta.
\end{eqnarray}
Here $a$ is the rotation parameter characterising the ergo-sphere of the rotating black hole.
\subsection{Phase structure}
Extended thermodynamics of rotating AdS black holes has been studied extensively in the literature \cite{Dolan119,45}. In order to set up the thermodynamics, let us note that the entropy (in the presence of non-zero rotation parameter \(a\)) is given as,
\begin{equation}
  S = \pi (r_+^2 + a^2)
\end{equation} where \(r_+\) is the horizon radius of the black hole. Now, the temperature and thermodynamic volume can be computed straightforwardly by taking derivatives of the ADM mass of the spacetime. They respectively read \cite{Dolan:2011xt},
\begin{equation}\label{T}
T= \frac{S^2 \left(64 P^2 S^2+32 P S+3\right)-12 \pi ^2 J^2}{4
   \sqrt{\pi } S^{3/2} \sqrt{(8 P S+3) \left(12 \pi ^2 J^2+S^2
   (8 P S+3)\right)}}
\end{equation}
and,
\begin{figure}[t]
\begin{center}
\includegraphics[width=5.0in]{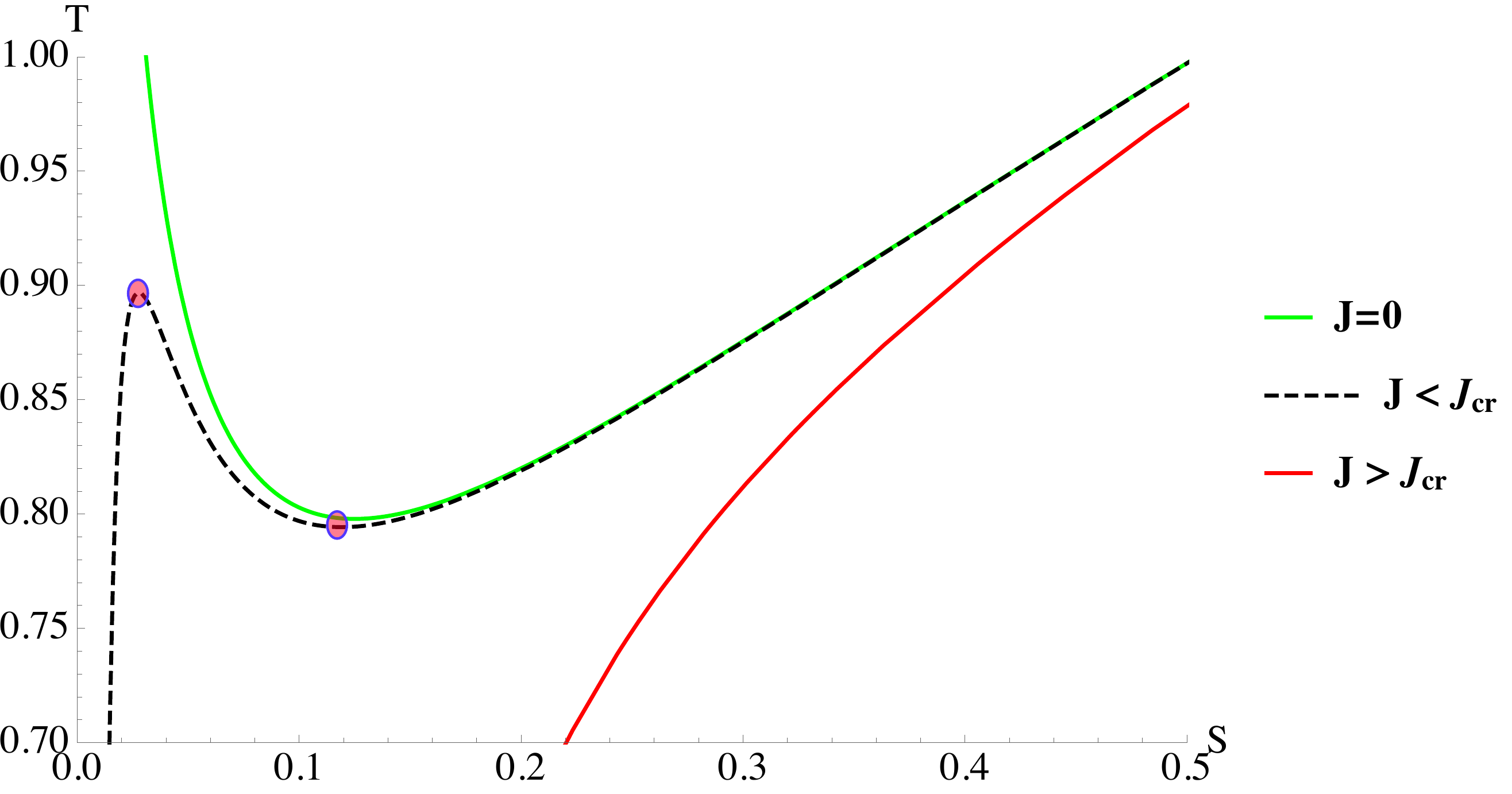}
\caption{\footnotesize Temperature vs entropy for rotating black holes in AdS for $P=1$ and various values of angular momentum. The value of critical angular momentum for the above choice of parameters is $J_{\rm cr}=0.0028567$. At a given temperature, depending on the value of \(J\), there can be up to three branches of black hole solutions. For the $J=0$ case (green curve), there are the familiar small and large black hole branches. For $J<J_{\rm cr}$ (black dashed curve, \(J=0.0021\)), there is a small branch, an intermediate branch and a large branch. The two red dots at which \(C_P\) diverges denote the beginning and end of the unstable intermediate branch. For $J > J_{\rm cr}$ (red curve, \(J=0.02\)), at any temperature, there is always a single large black hole branch. 
 } 
\label{T_S}
\end{center}
\end{figure}
\begin{equation}
V=\frac{4 \sqrt{S} \left(6 \pi ^2 J^2+S^2 (8 P S+3)\right)}{3
   \sqrt{\pi } \sqrt{(8 P S+3) \left(12 \pi ^2 J^2+S^2 (8 P
   S+3)\right)}}
\end{equation} 
where, \(J\) is the angular momentum parameter and is kept fixed. In extended thermodynamics, the ADM mass is thermodynamically equivalent to the enthalpy and not the internal energy, i.e. \(M := H\) and we have the specific heats \(C_{P} := T \partial S/\partial T |_{P}\) and \(C_{V} := T \partial S/\partial T |_{V}\).  The specific heats \(C_{P}\) and \(C_{V}\) are computed to be, 
\begin{equation}
C_{P}= \frac{\partial H}{\partial T}\Big|_{P} = \frac{2 S (8 P S+3) \left(12 \pi ^2 J^2+S^2 (8 P S+3)\right)
   \left(S^2 \left(64 P^2 S^2+32 P S+3\right)-12 \pi ^2
   J^2\right)}{24 \pi ^2 J^2 S^2 (8 P S+3)^2 (16 P S+3)+144
   \pi ^4 J^4 (32 P S+9)+S^4 (8 P S-1) (8 P S+3)^3}
\end{equation}
and,
\begin{equation}\label{CV}
C_{V}= \frac{\partial E}{\partial T}\Big|_{V}  = \frac{24 \pi ^4 J^4 S \left(S^2 \left(64 P^2 S^2+32 P
   S+3\right)-12 \pi ^2 J^2\right)}{48 \pi ^4 J^4 S^2 \left(64
   P^2 S^2+64 P S+15\right)+24 \pi ^2 J^2 S^4 (8 P S+3)^3+576
   \pi ^6 J^6+S^6 (8 P S+3)^4}
\end{equation} where, \(E = H - PV\). 
Isothermal compressibility is \cite{Dolan119},
\begin{equation}
\kappa_T=\frac{48 S \left(\frac{48 \pi ^4 J^4 (8 P S+3) (8 P
   S+5)}{S^4}+\frac{24 \pi ^2 J^2 (8 P S+3)^3}{S^2}+\frac{576
   \pi ^6 J^6}{S^6}+(8 P S+3)^4\right)}{\left(\frac{12 \pi ^2
   J^2}{S^2}+16 P S+6\right) \left(\frac{144 \pi ^4 J^4 (32 P
   S+9)}{S^4}+\frac{24 \pi ^2 J^2 (8 P S+3)^2 (16 P
   S+3)}{S^2}+(8 P S-1) (8 P S+3)^3\right)}.
\end{equation}
Figure-(\ref{T_S}) shows the temperature as a function of entropy (horizon radius) for different values of angular momentum. In the current (fixed \(J\)) ensemble, there exists a critical point and the phase structure is quite similar to the case of charged black holes in AdS studied in \cite{Chamblin}. The critical values in principle can be determined from the combined conditions,
\begin{equation} \label{dTdS}
  \bigg(\frac{\partial T}{\partial S}\bigg)_{J,P}= \bigg(\frac{\partial^2 T}{\partial S^2}\bigg)_{J,P} = 0 \, .
\end{equation}
We can solve these numerically~\cite{Wei:2015ana} and obtain the critical value quoted in figure-(\ref{T_S}), as an analytic expression is not obtainable, unless special limits are taken~\cite{45}. As seen from figure-(\ref{T_S}), for $J=0$, there are two branches, for $J<J_{\rm cr}$, there are three branches and for $J>J_{\rm cr}$ there is a single branch of black hole solutions. 
For instance for $J=0$, there are the usual small and large black hole branches, where, the former is unstable and the latter is stable, respectively \cite{Hawking:1982dh}. For $J<J_{\rm cr}$, there is a small branch, an intermediate branch and a large branch. The small and large black hole branches are locally stable whereas, the intermediate branch is unstable. For $J>J_{\rm cr}$, at any temperature, there is always a single large black hole branch. Stability of all the aforementioned branches of solutions is known from the behaviour of \(C_{P}\).  For the choice of parameters given in figure-(\ref{T_S}), the specific heat \(C_{P}\) diverges at two places (denoted by red dots), namely $S=0.0425$ and $S=0.112$, which are equilibrium points indicating new phases of the system. When $J=J_{\rm cr}$, the two points coalesce and \(C_{P}\) diverges at a single point $S=0.082$ (not shown in plots). We will later see in figure-(\ref{E_vs_S_J_Jcr}) that the fluctuations also diverge at this point. \\

\subsection{Thermodynamic fluctuations}
Before we compute thermodynamic fluctuations (the covariances) or the resulting entropy corrections, we should remark on the validity of our results. Firstly, the entropy corrections summarised in eqn (\ref{entropycorrections}) which can be computed from the expressions of the response functions are valid away from the critical point. At the critical point or close to it, thermodynamic fluctuations are large and it is not in general justified to consider the effect of fluctuations up to the second order (the Gaussian approximation). One straightforward way to see that this is indeed the case is to notice that as one approaches the critical point, \(\kappa_T\) diverges ensuring that \(D\) diverges too. Furthermore, below the critical temperature, \(\kappa_T\) may change sign leading to regions of both positive and negative compressibility. This feature is present even in the van der Waals fluid where there are unstable regions of negative compressibility which are dealt with using the Maxwell's construction. Thus, our corrections are valid sufficiently away from (above) the critical point. \\

\noindent
A second important aspect which should be emphasised upon is that our treatments regard black holes as thermodynamic systems within the realm of classical equilibrium thermodynamics. Thus, our logarithmic corrections may not provide an accurate picture near/at the extremal point where there are complicated fluctuation phenomena which are quantum mechanical in origin (see for example, \cite{extremal1,extremal2}). In such a limit, only a quantum gravity description is expected to provide consistent results. \\

\noindent
Consider the variance in energy, i.e. \((\Delta E)^2\) defined in eqn (\ref{EE}). Using the expressions of the response functions of the Kerr-AdS black hole in four dimensions, the expression for the relative squared fluctuations in energy is computed to be, 
\begin{equation}\label{EEKerr}
\frac{(\Delta E)^2}{E^2} = \frac{A_1 \times A_2}{ A_3 \times A_4}
\end{equation}
where,
\begin{equation*}
A_1 = S^4 (8 P S+3)^2 \left(S^2 \left(64 P^2 S^2+32 P S+3\right)-12 \pi ^2 J^2\right),
\end{equation*}
\begin{equation*}
A_2 = \left(48 \pi ^4 J^4 S^2 \left(64 P^2 S^2+16 P S-3\right)-12 \pi ^2 J^2 S^4 \left(64
   P^2 S^2+32 P S+3\right)+576 \pi ^6 J^6+S^6 (8 P S+3)^2\right),
\end{equation*}
\begin{equation*}
A_3 = 2 \left(24 \pi ^2 J^2
   S^2 (8 P S+3)^2 (16 P S+3)+144 \pi ^4 J^4 (32 P S+9)+S^4 (8 P S-1) (8 P S+3)^3\right),
\end{equation*}
\begin{multline*}
A_4 =  \Bigg(-4 \pi ^2 J^2 \bigg(3 \pi ^2 \sqrt{\frac{J^4 S}{(8 P S+3) \left(12 \pi ^2
   J^2+S^2 (8 P S+3)\right)}} \times  \\ 
 \sqrt{(8 P S+3) \left(12 \pi ^2 J^2+S^2 (8 P
   S+3)\right)}  -4 P S^{7/2}-3 S^{5/2}\bigg)+12 \pi ^4 J^4 \sqrt{S}+S^{9/2} (8 P
   S+3)\Bigg)^2.
\end{multline*}
The thermodynamic limit, i.e. $N \rightarrow \infty$ in the present case is equivalent to taking $S \rightarrow \infty$. This can be motivated as follows. In natural units, where \(G = l_P^2\), the black hole entropy \(S\) given by the Bekenstein-Hawking formula reads: \(S = A/4 l_P^2\). Thus, in a sense, if \(l_P\) is understood to be the microscopic length scale, then the entropy is proportional to the number of Planck area (\(l_P^2\)) pixels occupying the area \(A\) of the black hole. The microscopic degrees of freedom of the horizon are therefore proportional to \(S\) which counts the number of Planck area pixels occupying the total area, i.e. \(S \propto N\) (see for example \cite{dof1,dof2}). Henceforth, the thermodynamic limit may be motivated to be \(S \rightarrow \infty\) wherein the black hole carries a large entropy. \\
\begin{figure}[t]
\begin{center}
\includegraphics[width=5.0in]{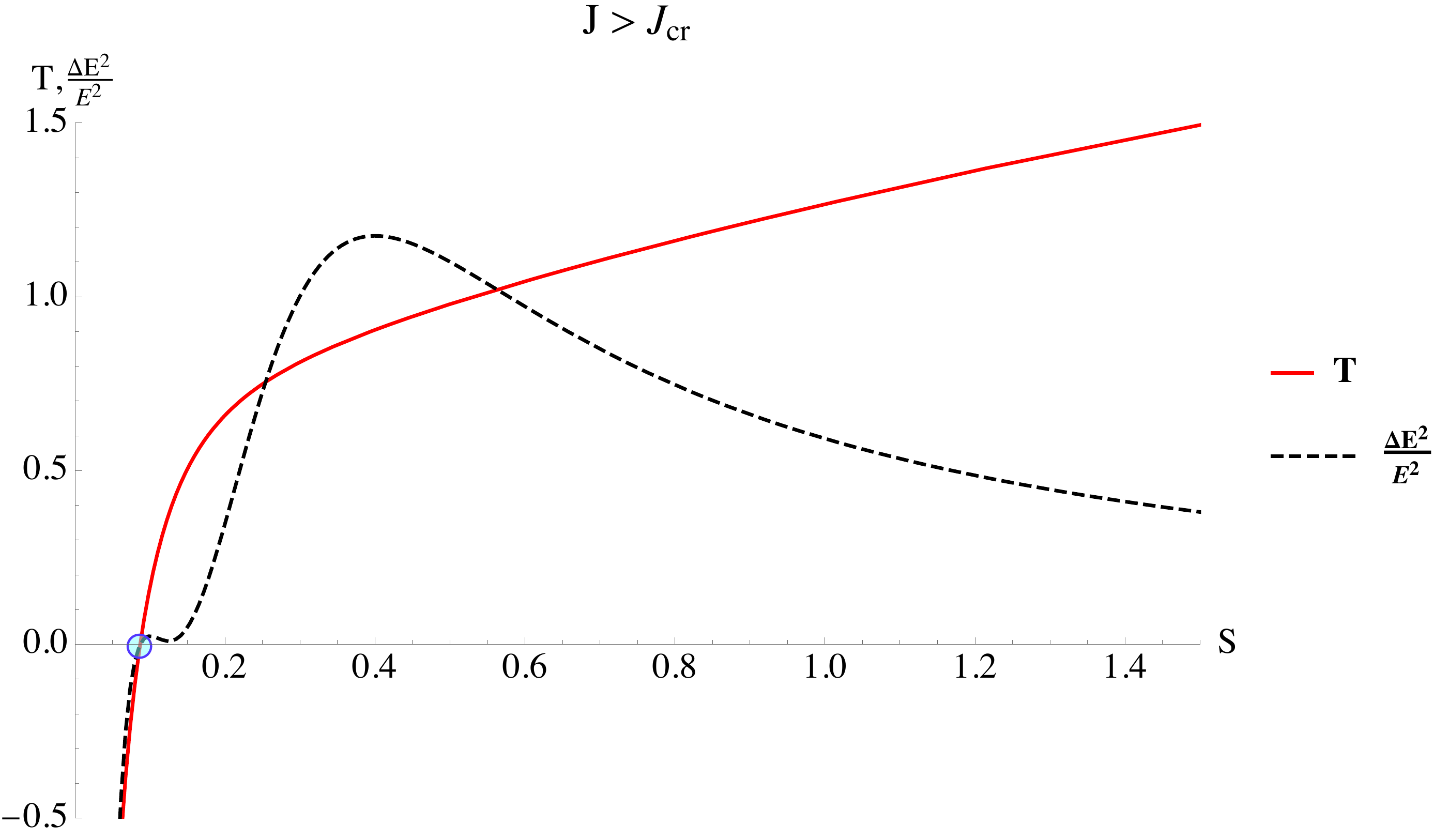}
\caption{\footnotesize Temperature and relative squared fluctuations in energy plotted for $J=0.02$ (\(J>J_{\rm cr}\)). The relative fluctuations rise from zero from the extremal point (blue dot, \(T=0\)), then go through a maximum and slowly go to zero at large temperatures. The maximum seen is an indication that relative fluctuations rise significantly at an intermediate point. 
 } 
\label{TEJ}
\end{center}
\end{figure}

\begin{figure}[t]
\begin{center}
\includegraphics[width=5.2in]{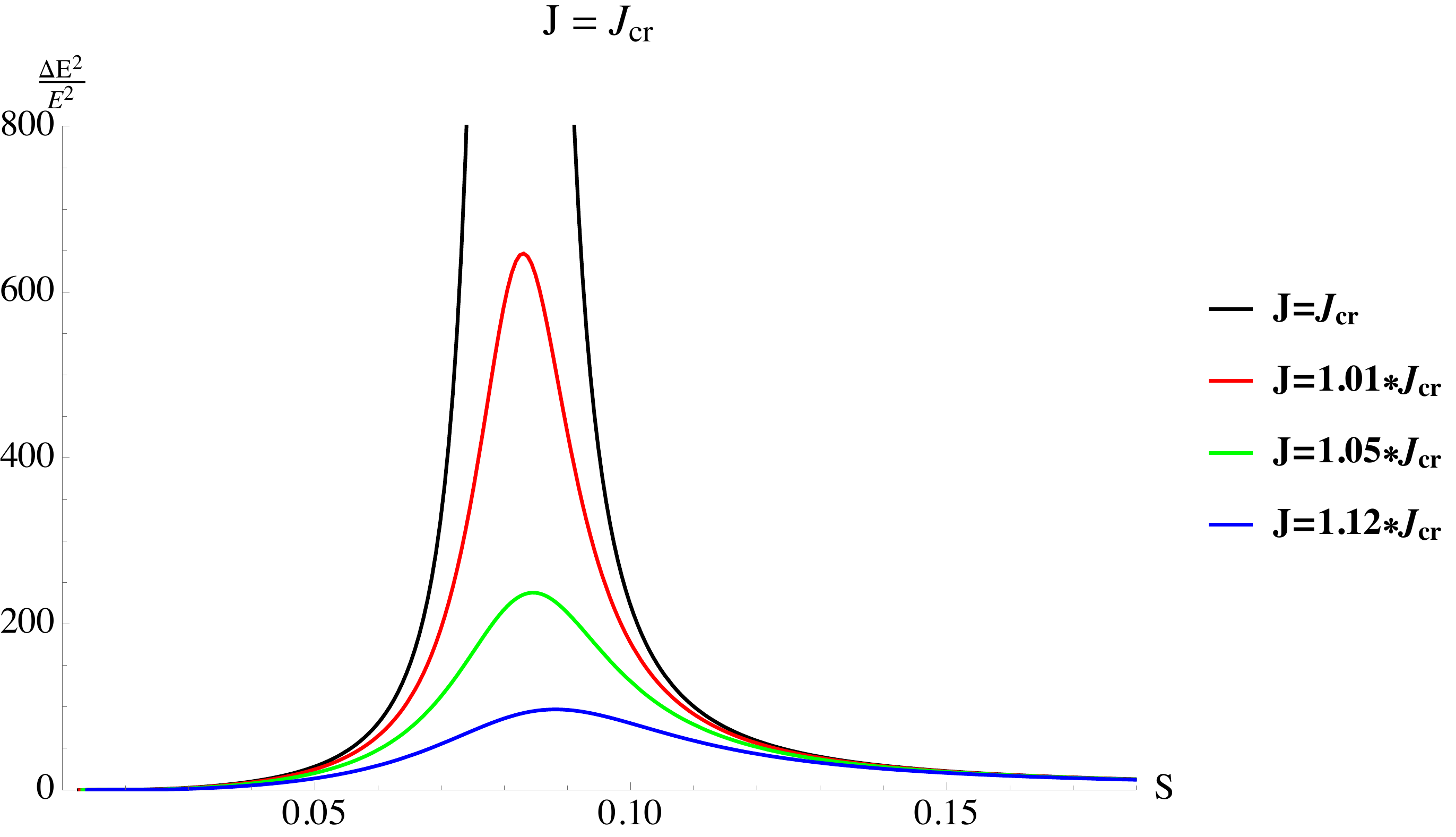}
\caption{\footnotesize Temperature and squared fluctuations in energy relative to equilibrium energy, plotted for $J\geq J_{\rm cr}$, where,  \(J_{\rm cr}=0.0028567\). At \(J=J_{\rm cr}\) (black curve), the fluctuations diverge at $S=0.082$ and it can be checked that $C_{P}$ also diverges at this point. The other curves from bottom to top show gradual decrease in \(J\) towards $J_{\rm cr}$.
 } 
\label{E_vs_S_J_Jcr}
\end{center}
\end{figure}

\begin{figure}[t]
\begin{center}
\includegraphics[width=6.1in]{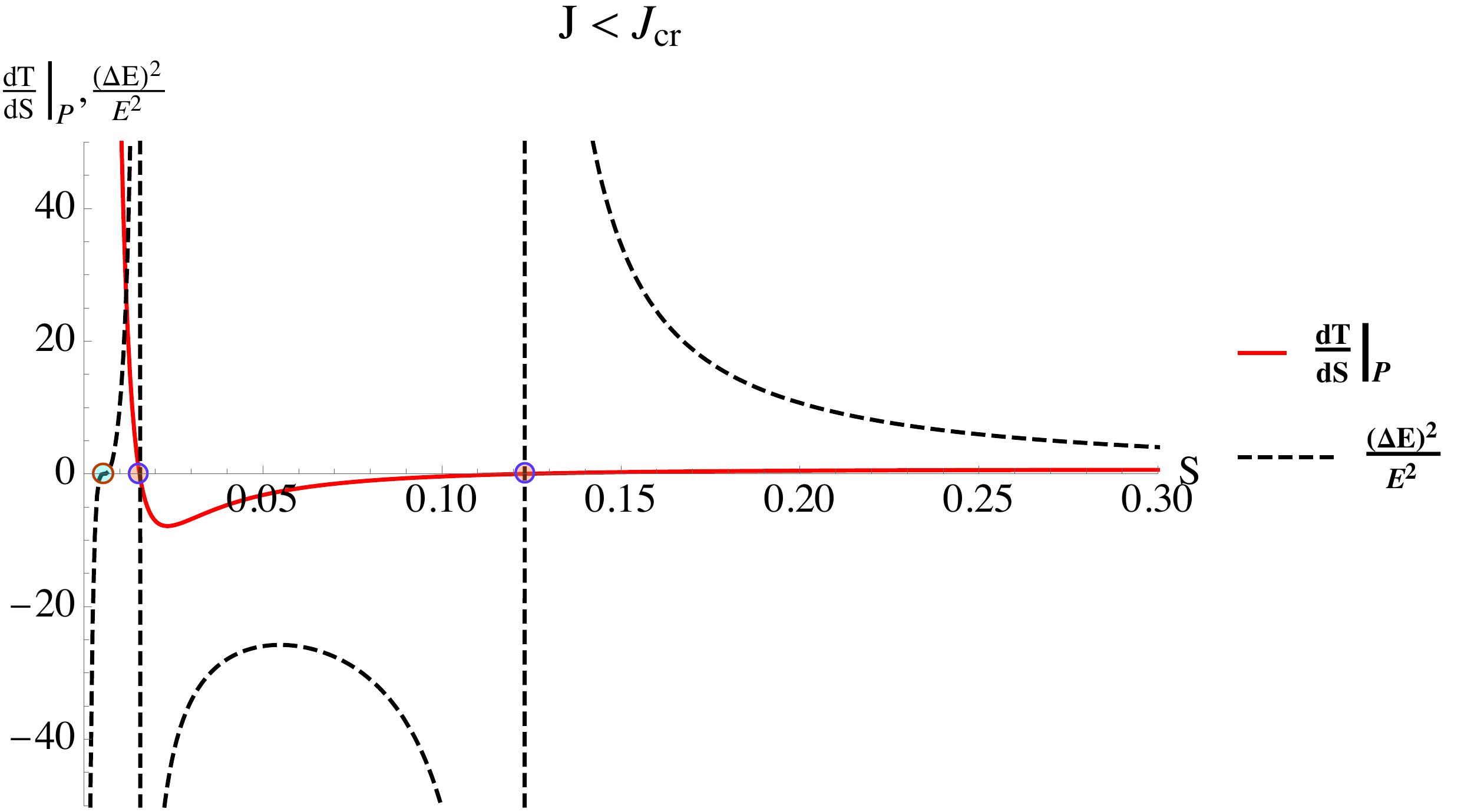}		
\caption{\footnotesize Slope of temperature showing stable regions (red curve) and squared fluctuations in energy relative to equilibrium energy (dashed curve) plotted for the case \(J = 0.0009 \,(< J_{\rm cr})\). The fluctuations start from zero (blue dot where \(T=0\)). The red curve is the slope of the temperature showing stability (at fixed \(P\)), and there are two red dots which show the points where the specific heat diverges. These are the same points marked in figure-(\ref{T_S}) too. We now see that the fluctuations in energy also diverge at these two points. In between these two points which come up at $S=0.0155$ and $S=0.123$, the fluctuations are negative, indicating instability. For large temperatures (here, at large entropy), the fluctuations of the single third branch (disconnected dashed curve to the right) go to zero and can be shown to go as \(1/T\). In summary, the fluctuations rise from zero, diverge at the phase transition point and then go to zero at high temperatures. } 
\label{dTdS_E_J_less_Jcr}
\end{center}
\end{figure}
\noindent
It follows from eqn (\ref{EEKerr}) that as \(S \rightarrow \infty\), the relative squared fluctuations of energy or \((\Delta E)^2/E^2\) scales with \(S\) as \((\Delta E)^2/E^2 \sim S^{-1}\). The relative (squared) fluctuations, i.e. \((\Delta E)^2/E^2\) have been plotted in figures-(\ref{TEJ}), (\ref{E_vs_S_J_Jcr}) and (\ref{dTdS_E_J_less_Jcr}) and clearly, they fall off rapidly as \(S\) gets large. In these plots \(J_{\rm cr}\) is the critical value of angular momentum associated with the phase transition of the Kerr-AdS black hole. Although the generic features for large \(S\) are the same in both the plots, the small \(S\) behaviours are different. In particular, it is seen that for \(J < J_{\rm cr}\), the relative squared fluctuations diverge at \((\partial T/\partial S)_P = 0\). These are precisely the points at which second order response functions such as \(C_P\) and \(\kappa_T\) diverge. In both the cases, i.e. for \(J > J_{\rm cr}\) and \(J < J_{\rm cr}\), the energy fluctuations are found to go to zero at \(T = 0\). This is quite expected because these thermal fluctuations arise due to finite temperature effects. In figure-(\ref{E_vs_S_J_Jcr}), the relative squared fluctuations are plotted for various values of angular momentum \(J\). They are found to diverge at the critical point for \(J = J_{\rm cr}\).  This divergence signifies a breakdown of the Gaussian approximation and is known to occur near a second order critical point, where the correlation lengths are known to diverge, as the order parameter vanishes. Here, the order parameter can be taken to be the difference between the horizon radii of the first and third branches. Larger values of angular momentum imply smaller relative fluctuations. This is consistent with our expectation that fluctuations are large near criticality. \\

\noindent
The relative squared fluctuations in volume are given by, 
\begin{equation}\label{VVkerr}
\frac{(\Delta V)^2}{V^2} = \frac{ B_1 \times B_2}{ B_3 \times B_4}
\end{equation}
where, 
\begin{equation*}
B_1 = 9 \left(S^2 \left(64 P^2 S^2+32 P S+3\right)-12 \pi ^2 J^2\right),
\end{equation*}
\begin{equation*}
B_2 = \left(48 \pi ^4
   J^4 S^2 \left(64 P^2 S^2+64 P S+15\right)+24 \pi ^2 J^2 S^4 (8 P S+3)^3+576 \pi ^6
   J^6+S^6 (8 P S+3)^4\right),
\end{equation*}
\begin{equation*}
B_3 = 2 \left(6 \pi ^2 J^2+S^2 (8 P S+3)\right)^2,
\end{equation*}
\begin{equation*}
B_4 =  \left(24 \pi
   ^2 J^2 S^3 (8 P S+3)^2 (16 P S+3)+144 \pi ^4 J^4 S (32 P S+9)+S^5 (8 P S-1) (8 P
   S+3)^3\right).
\end{equation*}
It is straightforward to verify that the relative fluctuations in thermodynamic volume follow the same qualitative trends as those of energy. At large \(S\), eqn (\ref{VVkerr}) implies that \((\Delta V)^2/V^2 \sim S^{-1}\), i.e. relative fluctuations in volume scales with \(S\) in the same way as the relative fluctuations in energy. They are also found to diverge at the critical point, where the response functions, in particular \(\kappa_T\) diverges. This is simply showing that the system is in an unstable regime and is along expected lines~\cite{Chamblin}. Similar conclusions are obtained for \(\langle \Delta E \Delta V \rangle\). We do not discuss it further. \\

\noindent
An interesting aspect of these relative fluctuations is that for both energy and thermodynamic volume, they scale with \(S\) inversely for large \(S\). If one takes \(S \sim N\), where \(N\) is the number of degrees of freedom, both energy and volume fluctuations scale in the large \(S\) limit as, 
\begin{equation}
\frac{\Delta E}{E} \sim \frac{1}{\sqrt{N}}, \hspace{5mm} \frac{\Delta V}{V} \sim \frac{1}{\sqrt{N}}
\end{equation} in exactly the same way as eqns (\ref{evn}) for a typical extensive system. This is rather remarkable, because black hole thermodynamics is not extensive, i.e. the black hole mass is not a homogenous function of its arguments\footnote{It is rather quasi-homogenous, with a scaling dictated by the Smarr formula \cite{Smarr} (see also \cite{Kastor:2009wy}).}. Thus, although in general \((\Delta E)/E\) may not scale so simply with \(N\), when the limit of large entropy is taken, a scaling analogous to that of extensive systems is obtained and the relative fluctuations fall off for \(N \rightarrow \infty\). 

\subsection{Corrections to black hole entropy}
We have discussed thermodynamic fluctuations and stability of Kerr-AdS black holes in the canonical and isothermal-isobaric ensembles. The latter describes the system more naturally, with the former arising from a Legendre transform. We can now present the logarithmic corrections to the entropy of these black holes in both canonical and isothermal-isobaric ensembles. With all the exact expressions of the response functions, the lowest order entropy corrections in the isothermal-isobaric ensemble can be computed directly from \(D\) which reads,
\begin{equation} \label{Dkerr}
D= \frac{C_1}{C_2 C_3}
\end{equation}
where
\begin{equation}
C_1= 12 \pi ^2 \left(J S^2 \left(64 P^2 S^2+32 P S+3\right)-12 \pi ^2
   J^3\right)^4 ,
\end{equation}
\begin{equation}
C_2= S^2 (8 P S+3)^2 \left(12 \pi ^2 J^2+S^2 (8 P S+3)\right)^2,
\end{equation}
\begin{equation}
C_3= 24 \pi ^2 J^2 S^2 (8 P S+3)^2 (16 P S+3)+144 \pi ^4 J^4 (32 P S+9)+S^4
   (8 P S-1) (8 P S+3)^3.
\end{equation}
Picking the leading term $1/S^2$ from eqn (\ref{Dkerr}), we get the following final expression for the log corrected entropy after neglecting constant terms and the terms of higher order,
\begin{equation}\label{Kerrentropycorrectionisothermalisobaric}
  \begin{split}
 \mathcal{S} = S_0 +  \ln S_0+ ({\rm other~terms~in~first~order}).
  \end{split}
\end{equation}
In the above equation, we have relabelled \(S \rightarrow S_0\) to signify that \(S_0\) is the equilibrium entropy given by the Bekenstein-Hawking area law whereas, \(\mathcal{S}\) is the microcanonical entropy of the system incorporating the logarithmic corrections owing to fluctuations about thermodynamic equilibrium. We should note that these corrections involve several other terms within the first order and they are not to be neglected. However, within this order, we find a term proportional to \(\ln S_0\) in conformity with earlier results on logarithmic corrections to black hole entropy. Comparing with eqn (\ref{k}), we find that \(k = -1\) describes these corrections. \\

\noindent
It is imperative to compare these logarithmic corrections with those obtained in the canonical ensemble. In the canonical ensemble, the volume of the system is fixed and is not allowed to fluctuate. The corresponding entropy corrections arise solely from thermal (energy) fluctuations and take the form given in eqn (\ref{correctedentropycanonical}) with \((\Delta E)^2 = T^2 C_V\). Using the expressions for \(T\) and \(C_V\) given in eqns (\ref{T}) and (\ref{CV}) respectively, the leading entropy corrections are of the form, 
\begin{equation}
  \begin{split}
 \mathcal{S}' = S_0 +  \ln S_0+ ({\rm other~terms~in~first~order})
  \end{split}
\end{equation} where we have put a prime to distinguish it from the one obtained in the isothermal-isobaric ensemble. Thus, we obtain the same leading order coefficient of the \(\ln S_0\) term as eqn (\ref{Kerrentropycorrectionisothermalisobaric}), i.e. \(k = -1\). We should however note that although the leading coefficient matches, the remaining terms which are not made explicit do not match. This is rather expected since the fluctuation properties of two different ensembles do not agree \cite{bravetti}. 

\subsubsection{High temperature and slowly rotating limits}
The high temperature limit of the Kerr-AdS black hole is obtained by taking \(S\) to be large. This is most clearly seen from eqn (\ref{T}) which in the large \(S\) limit scales as \(T \sim \sqrt{S}\) thereby reproducing the blackbody-like scaling between entropy and temperature in two spatial dimensions, i.e. \(S \sim T^2\). In this limit, \(D\) scales with \(S\) in the following manner, 
\begin{equation}\label{largeTD}
D \approx {\rm constants} \times \frac{J^4 S^{16}}{S^{10} S^8} \sim J^4 S^{-2}
\end{equation} 
where the \(P\) dependence is not made explicit. Thus, the leading order logarithmic corrections to the entropy take the following form, 
\begin{equation}
\mathcal{S} = S_0 + \ln S_0 + \cdots .
\end{equation} Therefore, the leading order dependence, i.e. the coefficient \(k = - 1\) is unaltered in this limit. \\

\noindent
One may also consider the slowly rotating or small angular momentum limit, i.e. \(J \approx 0\). It is easy to see that \(J \rightarrow 0\) (equivalently rotation parameter, \(a \rightarrow 0\)) is not smooth because it renders \(C_V = 0\) for the black hole. However, in the small angular momentum limit, one may retain in eqn (\ref{CV}), the lowest order term scaling as \(C_V \sim J^4 S^{-5}\). It follows that in this slowly rotating limit, \(D\) scales with \(S\) as \(D \sim  J^4 S^{-2}\), i.e. in a manner identical to eqn (\ref{largeTD}) leading to the same coefficient \(k = -1\) appearing in the resulting entropy corrections.

\section{Conclusions}\label{conclusions}
For a thermodynamic system in contact with energy and volume reservoirs, there are energy and volume fluctuations. For black holes in AdS, the effect of thermal fluctuations on entropy was computed in earlier works~\cite{parthasarathi,Mukherji:2002de}. In the present work, we have analysed the corrections to Bekenstein-Hawking entropy of black holes in AdS, due to energy as well as volume fluctuations and have found that they are logarithmic. We first obtained the general form of such corrections to the entropy and later applied them to the case of four dimensional Kerr-AdS black holes. The leading correction term which is of the form: \(\ln S_0\) carries the same coefficient in both canonical and isothermal-isobaric ensembles. Here \(S_0\) is the black hole entropy given by the Bekenstein-Hawking formula. Quite remarkably, we have found that the coefficient appearing in the \(\ln S_0\) term is unaltered in the high temperature and small angular momentum limits. These corrections to the black hole entropy are novel and are different from the ones reported earlier in literature.\\

\noindent
An interesting case is the \(J = 0\) case (see appendix-(\ref{appendixb})), where \(C_V = 0\) due to the dependence between entropy and thermodynamic volume. Subsequently, \(D = T^3 V C_V \kappa_T = 0\) which from eqn (\ref{entropycorrections1}) implies that \(\Delta E \Delta V = \langle \Delta E \Delta V \rangle\). Thus, the covariance of energy and volume is simply the product of their standard deviations. This implies that the correlation coefficient defined as,  
\begin{equation}
r = \frac{\langle \Delta E \Delta V \rangle}{ \Delta E \Delta V} = +1
\end{equation} indicating that these variables are perfectly correlated. In other words, any fluctuation in energy must be accompanied by a fluctuation in volume and vice versa. This rather intuitive since \(C_V = 0\), implying that the degrees of freedom of the black hole cannot be thermally excited at fixed volume. In general, for \(J \neq 0\), the correlation coefficient is smaller than \(+1\) meaning that the variables \(E\) and \(V\) are not perfectly correlated. Nevertheless, there is a non-zero correlation between energy and volume fluctuations since in general \(\langle \Delta E \Delta V \rangle \neq 0\).\\

\noindent
There are several avenues for future work. It would be nice to generalise our results by considering fluctuations in other parameters of the system and studying more general fluctuations together with charge and angular momentum as well (see for example \cite{open}). It is also important to pursue a possible microscopic explanation for the logarithmic corrections obtained here, as well as look for holographic insights. Recently, a novel framework has been put forward wherein the Newton's constant \(G\) is treated as a thermodynamic variable along with the cosmological constant \cite{visser,mann,zhao}. Such a framework for black hole thermodynamics can be mapped directly to the boundary and has the advantage of being extensive \cite{visser,zhao}. It would therefore also be interesting to study logarithmic corrections in such a set up where one takes into account fluctuations of the central charge in addition to energy and volume fluctuations. Finally, one should bear in mind that matching the microscopic and macroscopic computations of entropy has been one of the key aims of various theories of quantum gravity. While most of the computations correspond to black holes which are BPS, there have been attempts to obtain these corrections to near extremal cases~\cite{Sen:2012dw}. Thus, for further validity as well as for a
more conclusive understanding of these novel corrections, an investigation of the one loop determinant of
Euclidean gravity in this framework should be performed \cite{Bhattacharyya,Liu}. 
We keep these issues for future work.

\section*{Acknowledgements}
A.G. gratefully acknowledges the financial support received from the Ministry of Education (M.o.E.), Government of India in the form of a Prime Minister's Research Fellowship (ID: 1200454). S.M. thanks Goutam Tripathy for helpful discussions. C.B. thanks the Science and Engineering Research Board (SERB, DST), Government of India, through MATRICS (Mathematical Research Impact Centric Support) grant no. MTR/2020/000135.

\appendix

\section{Computing the covariances}\label{appendixa}
\numberwithin{equation}{section}

The elements of the covariance matrix are defined as, 
\begin{equation}\label{Dija}
D_{ij} = \langle(\Delta y_i)(\Delta y_j)\rangle = \frac{\partial^2 S}{\partial x^i \partial x^j}\bigg|_{\rm Equilibrium}
\end{equation} where, \( (x^1,x^2) = (\beta, \beta P) \) and \( (y_1,y_2) = (E,V)\). The entropy here is defined as \(S = \ln \Delta + \beta E + \beta PV\) with the saddle point equations,
\begin{eqnarray}
\bigg(\frac{\partial \ln \Delta (\beta,\beta P)}{\partial \beta}\bigg) &=& - E, \\
\bigg(\frac{\partial \ln \Delta (\beta,\beta P)}{\partial (\beta P)}\bigg) &=& - V.
\end{eqnarray}
Therefore, 
\begin{eqnarray}
 (\Delta E)^2  = \bigg(\frac{\partial^2 S(\beta,\beta P)}{\partial \beta^2}\bigg)_{\beta_0, (\beta P)_0}  &=& - \bigg(\frac{\partial E}{\partial \beta}\bigg)_{\beta P}  , \label{EE2} \\
\langle (\Delta E)(\Delta V)\rangle = \bigg(\frac{\partial^2 S(\beta,\beta P)}{\partial \beta \partial (\beta P)}\bigg)_{\beta_0, (\beta P)_0}  &=& - \bigg(\frac{\partial V}{\partial \beta}\bigg)_{\beta P} . \label{EV2} \\
\end{eqnarray}
The derivatives are evaluated at thermodynamic equilibrium, i.e. at \(\beta_0\) and \((\beta P)_0\). In the above equations, the derivatives with respect \(\beta\) are taken at fixed \(\beta P\). We can straightforwardly convert these derivatives to those at fixed \(\beta\) or \(P\) giving, 
\begin{eqnarray}
\frac{\partial E}{\partial \beta}\bigg|_{\beta P} &=&  \frac{\partial E}{\partial \beta}\bigg|_{P} - \frac{P}{\beta} \frac{\partial E}{\partial P}\bigg|_{\beta} \\
\end{eqnarray} and similarly for \(V\). Thus, eqns (\ref{EE2}) and (\ref{EV2}) can be re-written as, 
\begin{eqnarray}
 (\Delta E)^2 &=& -  \frac{\partial E}{\partial \beta}\bigg|_{P} + \frac{P}{\beta} \frac{\partial E}{\partial P}\bigg|_{\beta} ,\\
\langle (\Delta E)(\Delta V)\rangle &=&  -\frac{\partial V}{\partial \beta}\bigg|_{P} + \frac{P}{\beta} \frac{\partial V}{\partial P}\bigg|_{\beta}. \\
\end{eqnarray} which can be re-arranged to give the final expressions for the covariances \((\Delta E)^2\) and \(\langle(\Delta E)(\Delta V)\rangle\) given in eqns (\ref{EE}) and (\ref{EV}) respectively. From eqn (\ref{Dija}), \((\Delta V)^2\) is given by, 
\begin{eqnarray}
(\Delta V)^2  = \bigg(\frac{\partial^2 S(\beta,\beta P)}{\partial (\beta P)^2}\bigg)_{\beta_0, (\beta P)_0} = - \frac{\partial V}{\partial (\beta P)}\bigg|_{\beta}.  \label{VV3}   
\end{eqnarray}
Since the derivative on the left hand side involves a fixed temperature, we may write, 
\begin{equation}
(\Delta V)^2  = - \frac{1}{\beta} \frac{\partial V}{\partial P}\bigg|_{\beta} .
\end{equation} The above expression straightforwardly gives the expression for \( (\Delta V)^2 \) given in eqn (\ref{VV}).

\section{Spherically symmetric black holes}\label{appendixb}
\numberwithin{equation}{section}
In section-(\ref{bh}), we considered rotating black holes in AdS spacetimes and have analysed their thermodynamic fluctuations as well as obtained the entropy corrections due these fluctuations. Although individual covariances, i.e. \((\Delta E)^2\), \((\Delta V)^2\) and \(\langle \Delta E \Delta V \rangle \) can be obtained straightforwardly for the \(J = 0\) case corresponding to the spherically symmetric Schwarzschild-AdS black hole, the covariance matrix becomes singular. This is due to the fact that the Schwarzschild-AdS spacetime saturates the reverse isoperimetric inequality \cite{Cvetic:2010jb} thereby leading to a dependence between entropy \(S\) and thermodynamic volume \(V\). Thus, the specific heat at constant volume becomes, 
\begin{equation}
C_V = T \frac{\partial S}{\partial T}\bigg|_V = 0.
\end{equation} Since \(D\) is proportional to \(C_V\), it also vanishes making these entropy corrections invalid. This holds for all spherically symmetric black holes including charged black holes which saturate the reverse isoperimetric inequality. \\
\begin{table*}[t]
\caption{Logarithmic entropy corrections for some spherically symmetric black holes}
\centering
\begin{tabular}{| c | c | c |}
\hline
\textbf{Black hole} & \textbf{Leading entropy corrections} & \textbf{Value of coefficient \(\mathbf{k}\)}    \\ \hline
Non-rotating BTZ      & \(\mathcal{S} = S_0 - \frac{5}{2} \ln S_0 - \ln P\)       &  5/2  \\ 
& & \\
Rotating BTZ     & \(\mathcal{S} = S_0 + \frac{3}{2} \ln S_0 + \cdots\)       &  -3/2  \\ 
& & \\
Schwarzschild black holes in AdS\(_d\)     &  \(\mathcal{S} = S_0 - \frac{d+2}{6(d-2)} \ln S_0 + \cdots\)                       & \((d+2)/(6d-12)\)  \\ 
& & \\
 Charged black holes in AdS\(_d\)     &  \(\mathcal{S} = S_0 + \frac{1}{d-2} \ln S_0 + \cdots\)                       & \(-1/(d-2)\)  \\ 
& & \\
Five dimensional Gauss-Bonnet-AdS     &  \(\mathcal{S} = S_0 - \frac{1}{2} \ln S_0 + \cdots\)                       & \(1/2\)  \\\hline
\end{tabular}\label{table1}
\end{table*}
\\
\noindent
Following  \cite{Wei:2019uqg,Wei:2019yvs} (see also \cite{open,SarkarCv}),  a possible resolution to this situation is as follows. We may consider spherically symmetric black holes to be limiting cases of their non-spherically symmetric counterparts. This limit is equivalent to taking \(C_V \rightarrow 0^+\). As such, one may as a limiting case, put \(C_V = \epsilon\) where \(\epsilon\) is an infinitesimal positive constant, of the order of \(k_B\). This ensures that \(D\) is non-zero which reads \(D = \epsilon T^3 V \kappa_T\) so that the leading logarithmic corrections to the entropy are given as, 
\begin{equation}
\mathcal{S} = S_0 - \frac{1}{2} \ln (T^3 V \kappa_T) - \ln \sqrt{\epsilon} + {\rm higher~order~terms}.
\end{equation}
Since \(\epsilon\) is a constant, one may choose a suitable renormalisation scheme, where this constant can be absorbed, thereby giving finite logarithmic corrections to the entropy of spherically symmetric black holes. We list the leading order corrections for several black holes with spherically symmetric spacetimes in table-(\ref{table1}). Here, \((\cdots)\) imply other terms which are rational functions of \(S_0\) and other parameters. 
Of course, the procedure used above to extract the finite contribution to the corrections is rather ad hoc and further investigation is needed.

\end{document}